\documentclass{aa}  

\usepackage{graphicx}
\usepackage{siunitx}
\usepackage[lofdepth,lotdepth]{subfig}
\usepackage{natbib}
\usepackage{txfonts}
\usepackage[colorlinks = true,
            linkcolor = blue,
            urlcolor  = blue,
            citecolor = blue,
            anchorcolor = blue]{hyperref}
\usepackage{etoolbox}
\newcommand{\ubold}{\fontseries{b}\selectfont}
\robustify\ubold

\begin{document}

   \title{Torsional-rotational spectrum of doubly-deuterated dimethyl ether (CH$_3$OCHD$_2$)}

   \subtitle{First ALMA detection in the interstellar medium}

   \author{C. Richard
          \inst{1}
          \and
          J. K. Jørgensen\inst{2}
          \and
          L. Margulès\inst{3}
          \and
          R. A. Motiyenko\inst{3}
          \and
          J.-C. Guillemin\inst{4}
          \and
          P. Groner\inst{5}
          }

   \institute{Laboratoire Interdisciplinaire Carnot de Bourgogne, UMR 6303 CNRS - Universit\'e Bourgogne Franche-Comt\'e, 9 Av. A. Savary, BP 47870, F-21078 Dijon Cedex, France\\
              \email{cyril.richard@u-bourgogne.fr}
        \and
            Niels Bohr institute, University of Copenhagen, {\O}ster Voldgade 5-7, DK~1350 Copenhagen K., Denmark
        \and
            Univ. Lille, CNRS, UMR 8523 - PhLAM - Physique des Lasers Atomes et Molécules, 59000 Lille, France
        \and
             Univ Rennes, Ecole Nationale Supérieure de Chimie de Rennes, CNRS, IRCR-UMR 6226, F-35000 Rennes, France
        \and
            Department of Chemistry, University of Missouri-Kansas City, Kansas City, MO 64110-2499, USA
             }


 
  \abstract
   {In 2013, we have published the first rotational analysis and detection of mono-deuterated dimethyl ether in the solar-type protostar IRAS 16293-2422 with the IRAM 30~m telescope. Dimethyl ether is one of the most abundant complex organic molecules (COMs) in star-forming regions and their D-to-H (D/H) ratios is important to understand its chemistry and trace the source history.}
   {We present the first analysis of doubly-deuterated dimethyl ether (methoxy-\emph{d}$_2$-methane, 1,1-dideuteromethylether) in its ground-vibrational state, based on an effective Hamiltonian for an asymmetric rotor molecules with internal rotors. The analysis covers the frequency range 0.15–1.5~THz.}
   {The laboratory rotational spectrum of this species was measured between 150 and 1500~GHz with the Lille's submillimeter spectrometer. For the astronomical detection, we used the Atacama Large Millimeter/submillimeter Array (ALMA) observations from the Protostellar Interferometric Line Survey, PILS.}
   {New sets of spectroscopic parameters have been determined by a least squares fit with the ERHAM code for both symmetric and asymmetric conformers. As for the mono-deuterated species, these parameters have permitted the first identification in space of both conformers of a doubly-deuterated dimethyl ether via detection near the B component of the Class 0 protostar IRAS 16293-2422.}
   {}

   \titlerunning{Torsion-rotational spectrum of 2D-DME}
   \authorrunning{C. Richard et al.}
   \maketitle
%

\section{Introduction}
By may 2021, around 220 molecules have been detected in the interstellar medium (ISM) or circumstellar shells\footnote{https://cdms.astro.uni-koeln.de/classic/molecules}. Complex organic molecules (COMs) are ubiquitous in the ISM, especially close to the forming low-mass stars where dust temperature is high enough to sublimate ice. With its 9 atoms, dimethyl ether (CH$_3$---O---CH$_3$, hereafter DME), first detected in the ISM by~\citet{snyder1974} is a large COM of relevance for astrochemistry and one of the main COMs present in these warm and dense inner regions of the envelopes of Class 0 protostars, so called hot corinos~\citep{ceccarelli2004}. With the high sensitivity offered by new observations from the Atacama Large Millimeter/submillimeter Array (ALMA) new opportunities are opened-up for characterising the abundances of these COMs systematically \citep[see, e.g.,][for a recent review]{jorgensen20}.

One of the characteristics of the molecules present in the warm gas in hot corinos are their high degrees of deuteration. Although, the cosmic deuterium abundance relative to hydrogen is $1.5-2.0\times 10^{-5}$ \citep[e.g.,][]{linsky2003,prodanovic10}, the abundances of deuterated COMs relative to their non-deuterated counterparts are found to be much higher up to $\approx 10$\% or more in some cases \citep[e.g.,][]{parise06deuterium,jorgensen18}. These enhancements are thought to be a result of the exothermic reaction ${\rm H}_3^+ + {\rm HD} \quad\leftrightarrows\quad {\rm H}_2{\rm D}^+ + {\rm H}_2+ \Delta E$ that in gas at low temperatures enhances H$_2$D$^+$ relative to H$_3^+$. This effect is even more pronounced when one considers the multiple-deuterated variants that for several species (e.g., D$_2$CO, \citealt{persson18}; CHD$_2$OHCHO, \citealt{manigand19}, D$_2$O, \citealt{jensen21}) are even further enhanced compared to the relative abundances of the mono- and non-deuterated species, possibly related to the build-up of the ice-mantles where these species are formed.

Dimethyl ether is an interesting target species to extend these studies due to its structure and high abundance. The detection of the mono-deuterated form of CH$_3$OCH$_3$ toward the Class~0 protostellar system IRAS~16293-2422 was first reported by \cite{richard2013} based on the detection of 20 lines of its symmetric and asymmetric conformers. Through ALMA observations \cite{jorgensen18} determined the D/H ratio for these conformers (corrected for the number of equivalent hydrogen atoms) of about 3\% toward one component in this source, IRAS~16293B. With the high overall column density of DME toward this source, its D/H ratio and potential enhancement of the doubly deuterated species, this would be a natural place to test new spectroscopic predictions for the multi-deuterated variants.

\begin{figure}[ht]
  \begin{center}
    \subfloat[]{
      \includegraphics[width=0.4\columnwidth]{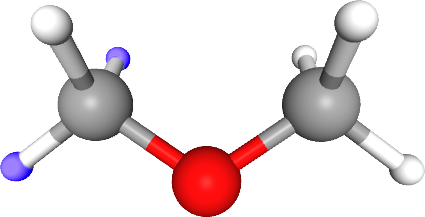}
      \label{sub:fig_a}
                }
                \LARGE +
    \subfloat[]{
      \includegraphics[width=0.4\columnwidth]{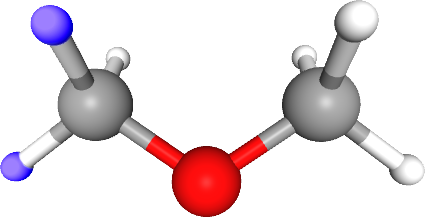}
      \label{sub:fig_b}
                }
    \newline
    \subfloat[]{
      \includegraphics[width=0.4\columnwidth]{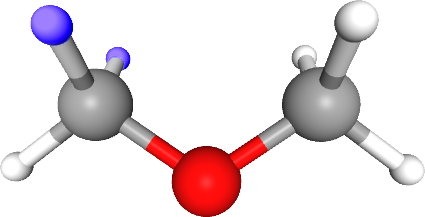}
      \label{sub:fig_c}
                }
    \caption{Representation of the doubly-deuterated DME with hydrogen atoms that are pictured in white and deuterium in blue. Figures~\ref{sub:fig_a} and \ref{sub:fig_b} show the two possible configurations for the asymmetric conformer while Fig.~\ref{sub:fig_c} illustrates the symmetric one.}
    \label{fig:configurations}
  \end{center}
\end{figure}

As we already have detailed in the analysis of the mono-deuterated species, DME is a near-prolate asymmetric top (Ray's asymmetry parameter $\kappa = -0.922$) with only a $b$-dipole moment component, $\mu_b = 1.302$~D~\citep{blukis1963}. The partial deuteration of one CH$_3$ group leads to two possible configurations. The Figure~\ref{fig:configurations} illustrates the molecule studied in this paper, the doubly-deuterated DME (2D-DME), also called methoxy-\emph{d}$_2$-methane in order to avoid confusion on the D atoms locations. The asymmetric conformation is identified when one of the deuterium is located on the C--O--C plane and the second in one of the other two locations. Therefore this conformer has two equivalent configurations (a) and (b) in Fig.~\ref{fig:configurations} with possible tunneling effect between them as detected for some lines of the mono-deuterated species \citep{richard2013}. When both deuterium atoms are outside of the C--O--C plane, the conformation is called symmetric, as it has a symmetry plane and thus belongs to $C_s$ symmetry point group. In this study, the two conformations of 2D-DME were considered as two independent asymmetric top molecules each with a single unsubstituted CH$_3$ internal rotor. The independence in the treatment of their rotational spectra consisted in ignoring the possible effects of tunneling between two equivalent configurations of the asymmetric conformation or between symmetric and asymmetric conformations.

This paper presents the first analysis of the 2D-DME from 0.15 to 1.5~THz (or 150 to 1500 GHz) and reports its detection toward the B component toward IRAS 16293-2422.

\section{Experimental details}
\subsection{Preparation of the doubly-deuterated DME (\texorpdfstring{methoxy-d$_2$-methane}{methoxy-d2-methane}, 1,1-dideuteromethylether).}
Potassium methoxide and \emph{p}-toluenesulfonyl chloride were purchased from Sigma-Aldrich. Methan-\emph{d}$_2$-ol was purchased from Eurisotop.

2D-DME DME was synthesized by the reaction of potassium methoxide with 4-methylbenzenesulfonic acid, methyl-\emph{d}$_2$ ester. This latter was prepared as previously reported for the trideutero derivative but using methan-\emph{d}$_2$-ol as alcohol \citep{yamamoto2016}. In a three-necked flask under nitrogen and connected to a trap immersed in a cold bath of dry ice were introduced potassium methoxide (5.6~g, 80~mmol) and dry DMSO (30~mL). The mixture was heated to 60°C and 4-methylbenzenesulfonic acid, methyl--\emph{d}$_2$ ester (8.2~g, 40~mmol) in dry DMSO (20~mL) was added dropwise. The doubly-deuterated DME was distilled off as it formed and condensed in the cold trap. At the end of the addition, the mixture was heated for 1 hour at 80°C. The trap was then closed with stopcocks. Yield: 1.0~g, 53\%. 

\subsection{NMR of the doubly-deuterated DME}
$^1$H NMR (CDCl$_3$, 400~MHz) $\delta$ 3.18 (quint, 1H, $^2$J$_{\text{HD}}$ = 1.6~Hz, CHD$_2$); 3.23 (s, 3H, CH$_3$). $^{13}$C NMR (CDCl$_3$, 100~MHz) $\delta$ 59.5 (quint, $^1$J$_{\text{CD}}$ = 21.6~Hz, $^1$J$_{\text{CH}}$ = 139.5~Hz (d), CHD$_2$); 60.1 (q, $^1$J$_{\text{CH}}$ = 140.4~Hz, CH$_3$).

\subsection{Lille THz spectrometer}
The absorption spectra were measured between 150 and 1500~GHz using the Lille spectrometer \cite{zakharenko2015}. The absorption cell was a stainless steel tube (6~cm in diameter, 220~cm in length). The measurements were performed at typical pressures of 35-60~Pa at room temperature. The different frequency ranges were covered with various active and passive frequency multipliers with the Agilent synthesizer (12.5--18.6~GHz) used as the primary signal source. To increase the sensitivity of the spectrometer, frequency modulation at 20.5~kHz of the reference source and lock-in detection are used. The demodulation of the detected signal may be performed either at $1f$ or $2f$, but $2f$ demodulation is preferred because of simpler presentation of observed spectrum in this case. Absorption signals were detected by an InSb liquid He-cooled bolometer (QMC Instruments Ltd.). Estimated uncertainties for measured line frequencies are 30~kHz, 50~kHz, and 100~kHz depending on the observed S/N ratio and the frequency range.

   \begin{figure*}
   \centering
   \includegraphics[width=0.8\textwidth]{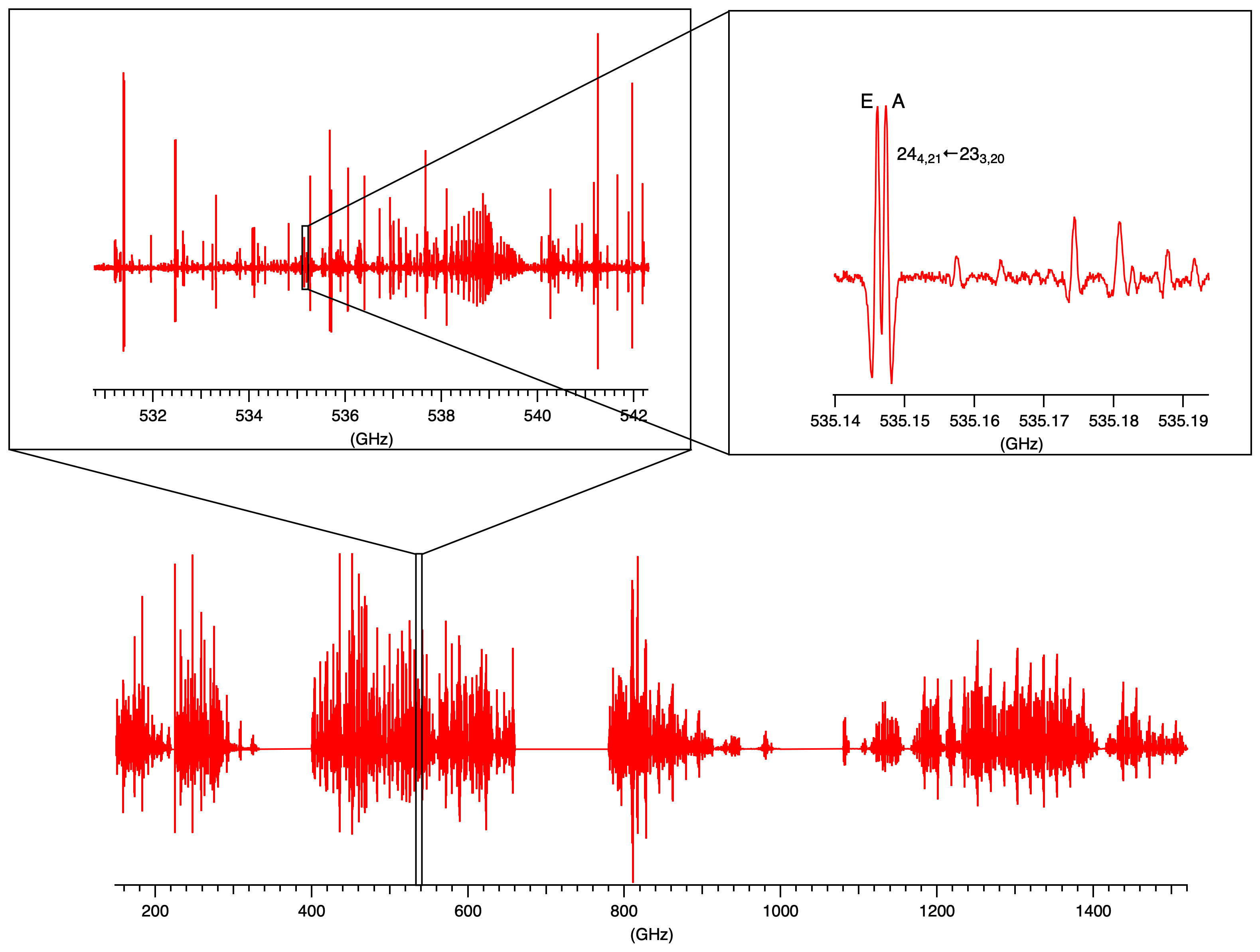}
      \caption{Spectrum of doubly-deuterated DME recorded in Lille. The lower panel shows the overall spectrum with a frequency range up to 1.5~THz. A zoom to some analyzed blended rotationnal transitions of the symmetric conformer is shown at the upper right.}
         \label{fig:spectrum}
   \end{figure*}

\section{Spectral analysis}\label{sec:analysis}

Similarly to the mono-deuterated species, the spectrum of the 2D-DME shows a very dense structure which is caused by the lines of the ground and the lowest excited vibrational states of symmetric and asymmetric conformations. In addition, due to internal rotation of the unsubstituted methyl top, each rotational level of the two conformations is split into $A$ and $E$ symmetry components of $C_{3v}$ symmetry point group. Despite relatively high barrier to internal rotation $V_3 \approx 900$~cm$^{-1}$ (see Table~\ref{tab:parameters}), the $A-E$ splittings of rotational transitions were resolved for the majority of the ground state lines of the 2D-DME adding thus more complexity to the measured spectra. To treat the internal rotation, we used the ERHAM code~\citep{groner1997,groner2012}. The model used in the code refers to the so-called "combined axes methods". It combines the rho-axis system in which the molecular Hamiltonian is set up, and the principal axes system to which the Hamiltonian is transformed. In the rho-axis system, the molecular internal $z$ axis is set parallel to the $\rho$ vector whose coordinates are calculated using the following expression:

\begin{equation}
 \rho_g = \frac{\lambda_g I_\alpha}{I_g}, \; (g = x, y, z)
\end{equation}

where $\lambda_g$ are the direction cosines of the internal rotation axis of the top in the principal axis system, $I_g$ are the principal inertia moments and $I_\alpha$ is the inertia moment of the methyl top. The magnitude of the rho vector represents the coupling between internal rotation of the methyl top and overall rotation of the molecular frame. For the two conformations of 2D-DME, the rho values are relatively high (see Table~\ref{tab:parameters}). They represent an intermediate case of the torsion-rotation coupling in the line between $cis$ methyl formate, HCOOCH$_3$, $\rho = 0.08$ \citep{ilyushin2009}, and acetaldehyde, CH$_3$C(O)H, $\rho=0.3$ \citep{smirnov2014}. For the latter two, strong torsion-rotation coupling, and medium height barriers to internal rotation significantly complicated the analysis and required the inclusion of many higher order terms into Hamiltonian. In the present analysis of the 2D-DME, high barrier to internal rotation simplified the treatment, and in addition to pure rotational Hamiltonian only 5 terms were needed to describe torsional-rotational interaction at close to experimental accuracy. 

The symmetric conformation of 2D-DME has a symmetry plane and thus belongs to the $C_s$ symmetry point group. Therefore, in the ERHAM code, the symmetry parameters ISCD is set to $-1$ and $\alpha$, the angle between the rho vector and $ab$ principal axes plane, is set to $0$. On the other hand, the asymmetric conformation has a minimal symmetry and belongs to the $C_1$ group. Consequently, for the asymmetric conformation, ISCD=$+1$ and after various attempts to set a value to $\alpha$, we have decided to fix the parameter to 0 for the same reasons as discussed in the Section~4 of the mono-deuterated analysis \citep{richard2013}.

The analysis was started with a pattern recognition of a few $K_a=0$ and $K_a=1$ $^aR_{0,1}$ lines in order to produce a fit and a first prediction. Then, the fit was iteratively improved, releasing more parameters, by adding new identified lines. 

   \begin{figure}[ht]
   \centering
   \includegraphics[width=\columnwidth]{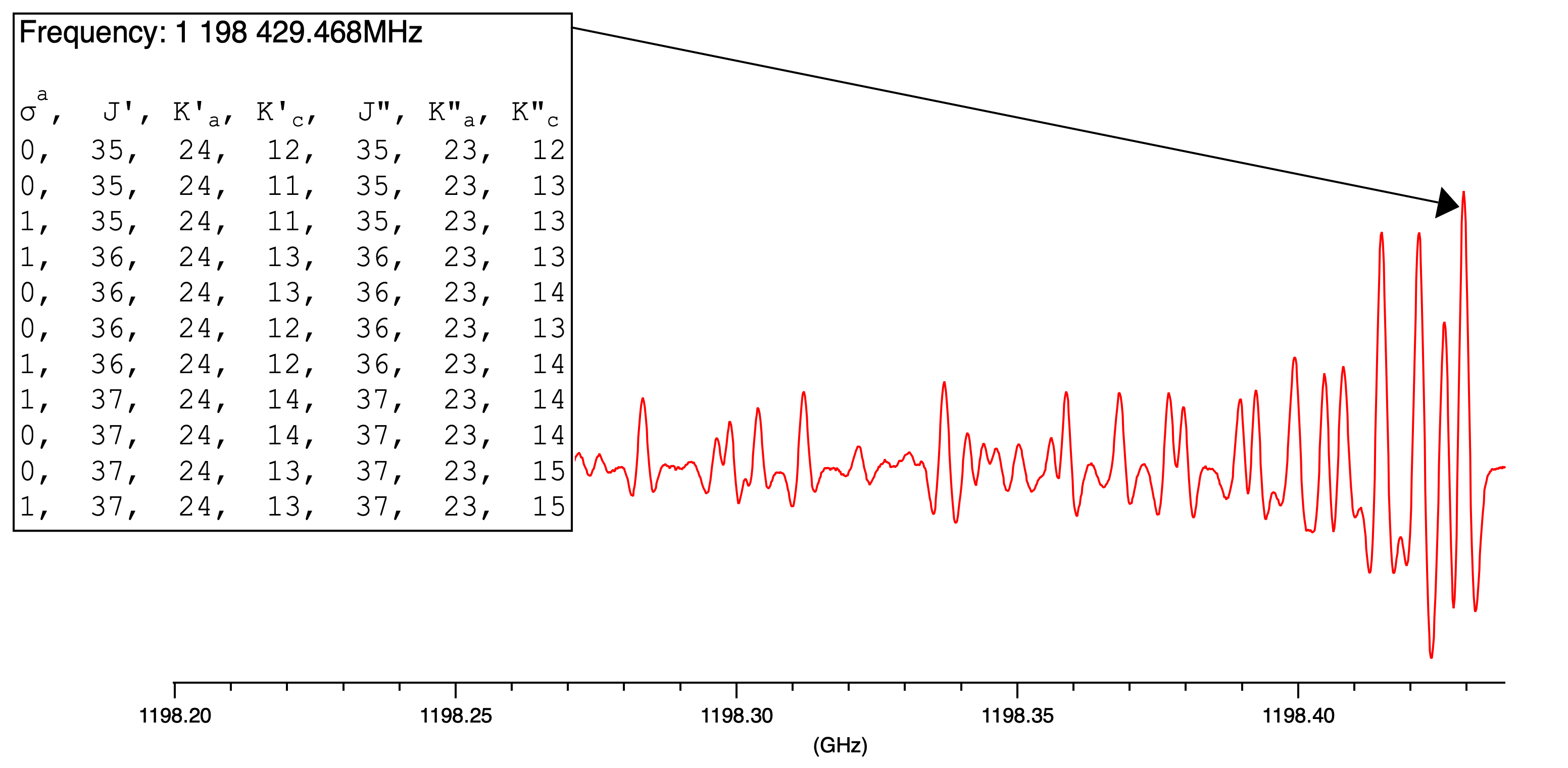}
      \caption{A spectrum zoom showing series with $K'=24$. All lines are extremely blended making the analysis very difficult and uninteresting. Therefore they are not been included in the fit.\\
      $^{(a)}$ symmetry number: $0(A)$, $1(E)$.
      }
         \label{fig:K_serie}
   \end{figure}

One difficulty encountered during the analysis was caused by several series of high $K_a$ $^bQ_{1,-1}$ transitions. An example is given in Fig.~\ref{fig:K_serie} for $K_a=24$ series. In this case, the spectral lines in the band head represent a congestion of many rotational transitions with different $J$ quantum numbers and different torsional symmetries. Such congestion has a strong influence on spectral line shape. Taking the spectral congestion and the second derivative spectrum into account, it leads to significantly reduced line frequency measurement accuracy, typically 3 to 10 times less than usual. In the weighted least-squares fit implemented in the ERHAM code, the weight of each frequency is calculated as reciprocal of the measurement accuracy squared. The influence of such highly congested lines would be 10 to 100 times less important. Therefore, we decided to remove these congested lines from the final fit and to include only the lines from such series that could be considered as isolated ones. 

\begin{table*}[ht]
 \caption{Spectroscopic constants of the ground-vibrational state of doubly-deuterated DME for the two diﬀerent conformers.}
\label{tab:parameters}
\begin{center}
\begin{tabular}{
	l
	S
	S
	}
	\hline\hline
\multicolumn{1}{c}{Parameters} & \multicolumn{1}{c}{\text{Symmetric}} & \multicolumn{1}{c}{\text{Asymmetric}} \\
 & \multicolumn{1}{c}{\text{conformer}} & \multicolumn{1}{c}{\text{conformer}} \\
\hline
\multicolumn{1}{l}{\emph{Pure Rotation}} \\
$A$ (MHz)   & 31492.87567(43) & 34199.90529(31) \\
$B$ (MHz)  & 9226.37143(19) & 8976.84575(12) \\
$C$ (MHz)  & 8256.93333(16) & 7984.12140(11)  \\
$\Delta_J$ (kHz)   &  8.18522(15) & 7.001791(72) \\
$\Delta_{JK}$ (kHz)& -6.03171(74) & -12.80820(49)  \\
$\Delta_K$ (kHz)& 167.2067(20) & 208.1287(14) \\
$\delta_J$ (kHz)& 1.541683(51) &  1.335158(25) \\
$\delta_K$ (kHz)& -31.8881(24) &  -4.1331(15) \\
$\Phi_J$ (Hz)   & .007359(31)  & .003928(15)  \\
$\Phi_{JK}$ (Hz)&  .1469(29)   & .0908(12) \\
$\Phi_{KJ}$ (Hz)& -2.5745(84)  & -2.8404(39)  \\
$\Phi_K$ (Hz)   & 5.8044(60)   & 6.8016(32) \\
$\phi_J$ (Hz)   & .003775(15)  & .001907(68) \\
$\phi_{JK}$ (Hz)& .50088(97)   & 0.10031(62) \\
$\phi_K$ (Hz)   & 1.779(83)    & 2.857(41) \\
\\
\multicolumn{1}{l}{\emph{Tunneling terms}} \\
$\rho$ & 0.17032(43) & 0.19123(14)  \\ 
$\beta$ (deg) & 10.935(47) &  8.718(19)  \\ 
$\alpha$ (deg) &  0.0\tablefootmark{a} & 0.0\tablefootmark{b} \\ 

$\varepsilon_{10}$ (MHz)    & -2.1814(73) & -2.4740(42) \\
$[A - (B + C)/2]_{q=1}$ (kHz) & .443(65) & 0.438(49)  \\
$[(B - C)/4]_{q=1}$ (kHz)    & .0626(92) & .0805(40)  \\
\\
$V_3\tablefootmark{c}$ (10$^{2}$cm$^{-1}$) & 9.014(30) & 9.044(15) \\
\\
Lines fitted\tablefootmark{d} & 1451 & 2080  \\
$J$(max), $K_a$(max) & \multicolumn{1}{c}{\text{88, 27}} & \multicolumn{1}{c}{\text{74, 25}} \\
$L_\text{worst}$\tablefootmark{e} & -4.4 & -4.8 \\
$n$\tablefootmark{f} & 20 & 20\\
$\sigma_\text{fit}$\tablefootmark{gf} (MHz)& 0.080  & 0.077  \\
$\sigma_\text{w}$\tablefootmark{h}& 1.09  & 1.01  \\

\hline
\end{tabular}
\tablefoot{Numbers in parentheses are one standard deviation in the same units as the last digit.\\
\tablefoottext{a}{By symmetry}
\tablefoottext{b}{Assumed, see Sect.~4 in \citet{richard2013}}
\tablefoottext{c}{Barrier height for rotation of the methyl group determined with the program BARRIER. See Sect.~\ref{sec:discussion} for uncertainties computation.}
\tablefoottext{d}{Number of distinct lines in the fit.}
\tablefoottext{e}{$(o. - c.)/$error of the poorest-fit line.}
\tablefoottext{f}{Number of free parameters used in the fit.}
\tablefoottext{g}{Standard deviation of the fit.}
\tablefoottext{h}{Weighted standard deviation of the fit.}
}
\end{center}
\end{table*}

A total of 1451 lines with $J_{max}=71$, $K_{a,max}=27$ and 2080 lines with $J_{max}=72$, $K_{a,max}=25$ were assigned respectively for symmetric and asymmetric conformers. For each conformer, a set of 15 pure rotational and 5 torsional-rotation parameters of the effective Hamiltonian implemented in the ERHAM code was needed to fit the observed molecular lines. The determined parameters are listed in Table~\ref{tab:parameters}. In the torsional-rotational part of the Hamiltonian, besides $\rho$, we determined the following parameters: $\beta$ the angle between rho vector and principal axis $a$, $\varepsilon_{10}$ the tunneling parameter representing the splitting between $A$ and $E$ symmetry sublevels, and two rotational constant tunneling parameters ($[A - (B + C)/2]_{q=1}$ and $[(B - C)/4]_{q=1}$).

\section{Discussion}\label{sec:discussion}
This study represents the first comprehensive characterization of the rotational spectrum of the two conformations of 2D-DME (CH$_3$OCHD$_2$) in the ground vibrational state, and in the frequency range up to 1.5 THz. Experimental measurements, partially given in appendices in Table~\ref{tab:exp_conf_s} and \ref{tab:exp_conf_a} are available in their entirety in electronic form at the Centre de Données astronomiques de Strasbourg (CDS). 

In the ERHAM code, the barrier to internal rotation is not calculated explicitly. Consequently, we used the program BARRIER which is derived from the program ASTOR described in \citet{groner1986} that determines the potential barrier for a single simple internal rotor from torsional transitions or splittings. The ground state energy splitting between the $A$ and $E$ symmetry torsional substates, $\Delta E(E-A)$, was combined with the derived value of the internal rotation constant $F$ to obtain an estimate of the potential barrier $V_3$ for both conformers. These values are direct output of ERHAM. In particular, the internal rotation constants $F$,  is obtained by ERHAM from $\rho$, $\beta$ and the rotational constants. The $F$ values determined in this study are 6.223 and $6.338$~cm$^{-1}$, respectively, for symmetric and asymmetric conformers. The determined $V_3$ values are given in Table~\ref{tab:parameters}. The estimated error of the $V_3$ barriers is derived with the assumption that the relative error of $V_3$ is the same as the relative errors of the $\Delta  E(E-A)$ energy difference and the tunneling coefficient $\varepsilon_{10}$. We found respectively an uncertainty of 3 and 1.5~cm$^{-1}$ for the symmetric and asymmetric conformers. The determined barrier heights agree well with the results of the study of monodeuterated DME \citep{richard2013}, and with earlier studies of parent DME \citep{durig1976, lovas1979}. 

Under Doppler limited spectral resolution, we didn't observe any additional splittings in the ground state rotational lines of the asymmetric conformation as it could be expected owing to possible tunneling between its two equivalent configurations, and as it was previously observed for monodeuterated DME \citep{richard2013}. It should be noted that for the latter, the tunneling splittings were barely resolved for a limited number of lines. The lack of additional tunneling splittings for the asymmetric conformer may be explained by the increased mass of the internal rotor which is composed of two deuteriums and one hydrogen for 2D-DME. Provided that the barrier height does not vary significantly under H to D isotopic substitution, the increased internal rotor mass reduces the tunneling probability and consequently the splitting between the two tunneling substates (compared to monodeuterated DME). In this case, one could also expect smaller and unresolvable splittings under the Doppler limit in the rotational spectra. The absence of resolved additional splittings as well as the weighted rms deviations of the fits close to 1 support our initial strategy to treat the two conformations independently as two asymmetric rotors each having a single methyl top.

\section{Predictions}
Thanks to the new set of spectroscopic parameters given in Table~\ref{tab:parameters} and using the ERHAM software in its ``ERHAM V16g-R3'' version (downloaded on the PROSPE website~\citep{kisiel2001}), we calculated predictions in the JPL catalog format~\citep{pickett1991} up to 1.5~THz.

Two short examples are provided in Tables~\ref{tab:pred_conf_s} and \ref{tab:pred_conf_a} from 150.1 GHz to 153.5 GHz. The complete tables are available through the CDS in the catalog format.

Numerical values of overall partition functions were computed at 9 different temperatures using the ERHAM output and are reported in the Table~\ref{tab:partition}. A good approximation of these values can be obtained from the simple rigid asymmetric rotor approximation \citep{townes1975} and is expressed in the following formula:
\begin{equation}\label{eq:partition_function}
 g_n \sum_{J}(2J+1)e^{-\frac{E_{\text{rot}}(J,K)}{kT}} =  g_n\sqrt{\frac{\pi}{ABC}\left( \frac{k}{h}\right)^3}T^{3/2}
\end{equation}

In equation~\ref{eq:partition_function}, the degeneracy factor $g_n$ is equal to 1 for the symmetric conformation and 2 for the asymmetric one, since the latter has two equivalent configurations and each energy level is split into two sublevels. We thus obtain 27.53433$T^{3/2}$ for the symmetric conformer and 54.48133$T^{3/2}$ for the asymmetric one. These values have also been listed in the Table~\ref{tab:partition} for comparison purposes. However, only the results from the fit, highlighted in bold, were used to derive column densities.

\begin{table*}[ht]
\caption{Rotational partition function for the symmetric and asymmetric conformers of doubly-deuterated DME in the ground vibrational state computed for nine different temperatures.}
\label{tab:partition}
\begin{center}
\begin{tabular}{
	l
	S[detect-weight,table-format = 2.3]
	S[detect-weight,table-format = 2.3]
	S[detect-weight,table-format = 2.3]
	S[detect-weight,table-format = 3.3]
	S[detect-weight,table-format = 5.3]
	S[detect-weight,table-format = 5.3]
	S[detect-weight,table-format = 5.3]
	S[detect-weight,table-format = 5.3]
	S[detect-weight,table-format = 5.3]
	}
\hline\hline 
Temperature & \multicolumn{1}{c}{2.275 K} & \multicolumn{1}{c}{5 K} & \multicolumn{1}{c}{9.375 K} & \multicolumn{1}{c}{18.75 K} & \multicolumn{1}{c}{37.5 K} & \multicolumn{1}{c}{75 K} & \multicolumn{1}{c}{150 K} & \multicolumn{1}{c}{225 K} & \multicolumn{1}{c}{300 K} \\
\hline
\multicolumn{10}{c}{Symmetric conformer} \\
Approx.\tablefootmark{a} & 94.481 & 307.843 & 790.372 & 2235.509 & 6322.974 & 17884.072 & 50583.794 & 92928.364 & 143072.576 \\
\textbf{Total} & \ubold 100.006 & \ubold 315.891 & \ubold 801.356 & \ubold 2251.224 & \ubold 6346.528 & \ubold 17925.034 & \ubold 50684.990 & \ubold 93140.580 & \ubold 143460.262\\
Ratio & 1.058 & 1.026 & 1.014 & 1.007 & 1.004 & 1.002 & 1.002 & 1.002 & 1.003 \\
\\
\multicolumn{10}{c}{Asymmetric conformer} \\
Approx.\tablefootmark{a} & 186.947 & 609.120 & 1563.884 & 4423.333 & 12511.074 & 35386.660 & 100088.590 & 183874.482 & 283093.284\\
\textbf{Total} & \ubold 197.606 & \ubold 624.644 & \ubold 1585.062 & \ubold 4453.594 & \ubold 12556.198 & \ubold 35463.956 & \ubold 100274.098 & \ubold 184257.132 & \ubold 283786.136\\
Ratio & 1.057 & 1.025 & 1.013 & 1.007 & 1.004 & 1.002 & 1.002 & 1.002 & 1.002 \\

\hline
\end{tabular}
\tablefoottext{a}{These values are computed with the rigid asymmetric rotor approximation}
\end{center}
\end{table*}

\section{\texorpdfstring{Astronomical search for CH$_3$OCHD$_2$}{Astronomical search for CH3OCHD2}}
As an application of the new spectroscopy we searched for the two conformers of CH$_3$OCHD$_2$ in Atacama Large Millimeter/submillimeter Array (ALMA) observations from the Protostellar Interferometric Line Survey, PILS \citep{jorgensen16}. The PILS program is an unbiased molecular line survey of the protostellar system IRAS~16293-2422 with ALMA (project id: 2013.1.00278.S; PI: Jes J{\o}rgensen). PILS covers a continuous frequency range from 329.1 to 362.9~GHz of the main atmospheric window in ALMA's Band~7. The spectral resolution over this full range is $\approx$~0.2~km~s$^{-1}$ and the sensitivity $\approx$~5~mJy~beam$^{-1}$~km~s$^{-1}$ with an angular resolution of 0.5$''$.

The advantage of looking for the deuterated species toward IRAS~16293-2422 is that it shows high D/H ratios for many organic molecules compared to, in particular, high-mass star forming regions otherwise targeted in studies of complex organic chemistry \citep[e.g.][and references therein]{jorgensen20}. For example, it was toward this source that a number of multiple deuterated species were identified for the first time \citep[e.g.][]{loinard00,parise02,parise03,manigand19} and recently the PILS survey has provided a comprehensive inventory of the deuterated content of the warm complex organic molecules close to the protostar \citep{jorgensen18,persson18,manigand20}.

To look for CH$_3$OCHD$_2$ we here targeted a position offset by 0.5$''$ to the southwest from the ``B'' component of the system (IRAS16293B) -- also the target of a number of the previous PILS papers. The advantage of searching at that position is that the lines are narrow ($\approx~$1~km~s$^{-1}$) and that the opacity of both the main lines of the dust continuum is limited. This enables easy assignments of individual transitions without problems with line overlaps and absorption.

At the targeted scales, the densities should be high enough for the excitation of the molecules to occur under Local Thermodynamic Equilibrium (LTE) conditions. We consequently used synthetic spectra calculated under this assumption to identify the lines of CH$_3$OCHD$_2$ within the PILS range and using them to constrain its column density. The synthetic spectrum method is preferred over, e.g., classical rotation diagrams, as it immediately makes it possible both to search for good matches in the data and to show if lines are predicted at frequencies where no emission is seen. In particular, to confirm the presence of rare species to demonstrate that the latter is not the case is equally important in finding good assigned matches. Also, this method makes it possible to predict the profiles for partially blended lines and also to look for overlaps with previously identified species.

Under the LTE assumption the line profiles are calculated assuming Gaussian line-profiles with constant LSR velocity offsets and line widths. Also, the extent of the source emission (and consequently beam filling factor) is kept constant. For these parameters we adopt the same values as for the single-deuterated variants and most other species in the PILS range -- a LSR velocity of 2.6~km~s$^{-1}$, a line width (FWHM) of 1~km~s$^{-1}$\ and source size of 0.5$''$ (assumed as a Gaussian distribution). The choices of the LSR velocity and line width can be tested directly when comparing to the observed spectra. The assumed source size mainly affects the results when optically thick lines are considered: in the case of optically thin emission and when species that thought to be co-existent in the gas it does not affect the derived relative abundances.

With these choices of fixed parameters what remains to be fitted are the column densities of the species and their excitation temperature. For the latter we adopt a temperature of 125~K of the non- and singly-deuterated variants of dimethylether, again under the assumption that the doubly-deuterated variants are present in the same gas. We then perform $\chi^2$-fitting to the reasonably isolated lines to constrain the column densities of the symmetic and asymmetric conformers separately.

\begin{figure*}[ht]
  \begin{center}
    \subfloat[]{
      \includegraphics[width=0.8\columnwidth]{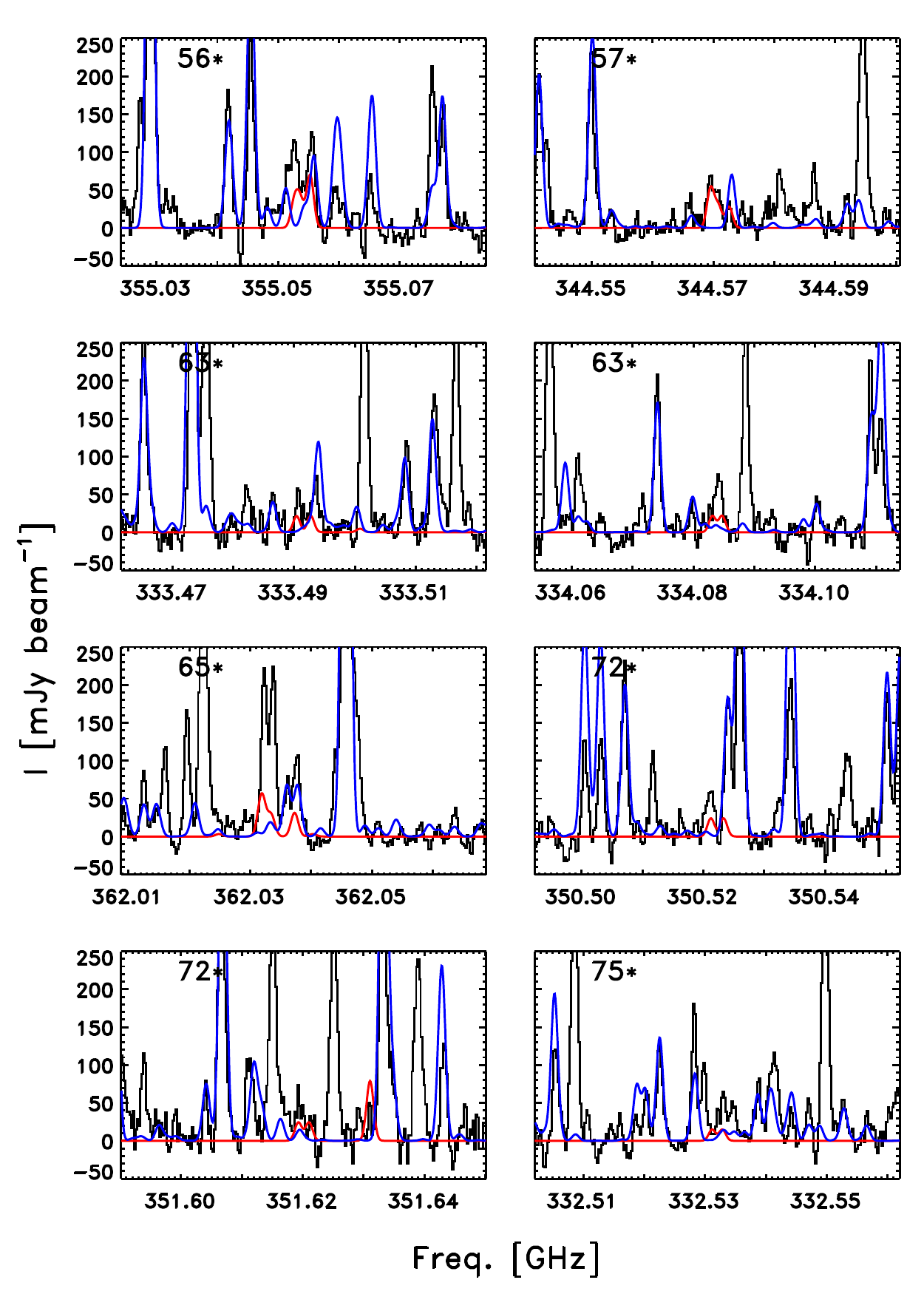}
      \label{sub:d-fullbeam_a}
                }
    \subfloat[]{
      \includegraphics[width=0.8\columnwidth]{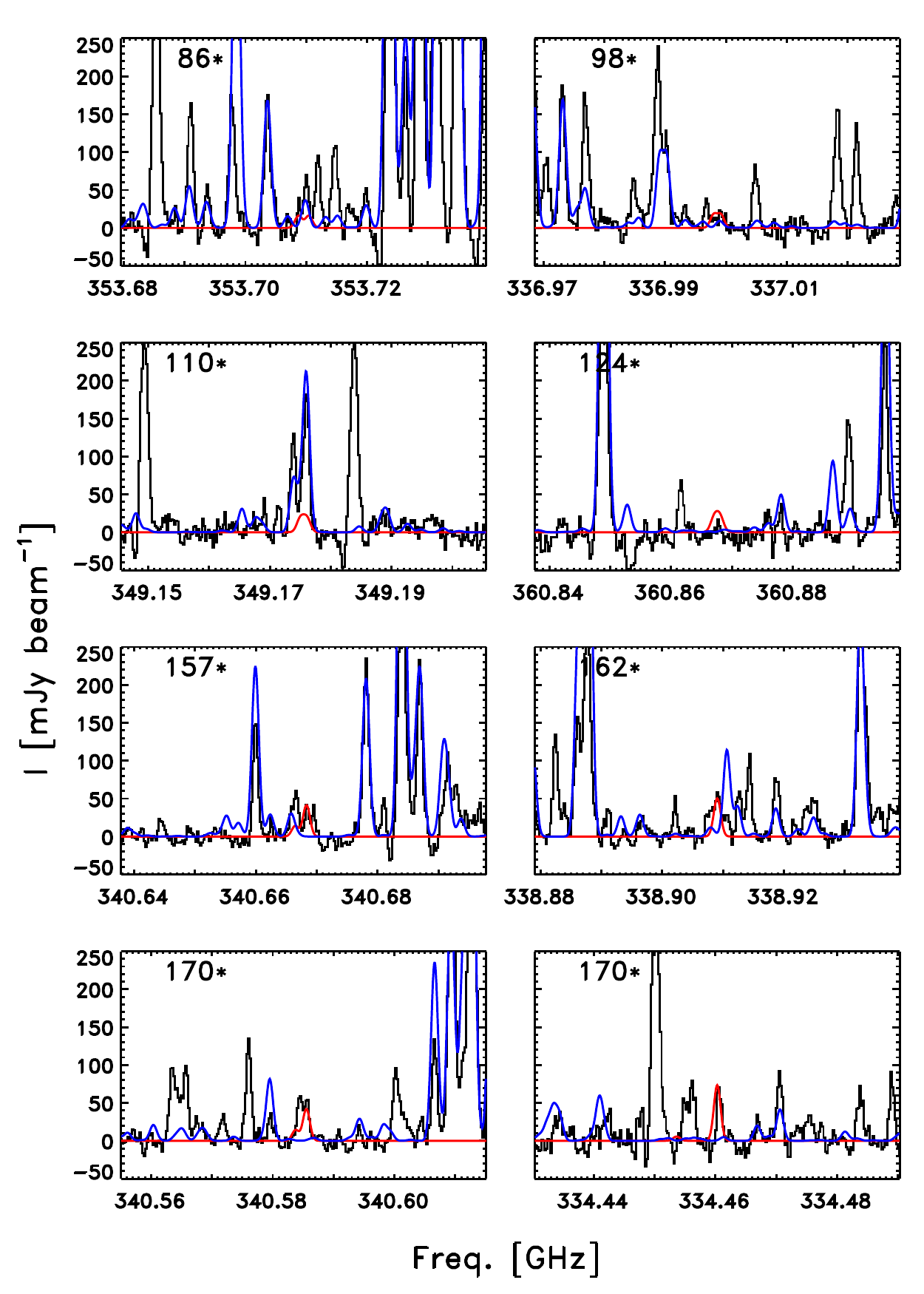}
      \label{sub:d-fullbeam_b}
                }
    \newline
    \subfloat[]{
      \includegraphics[width=0.8\columnwidth]{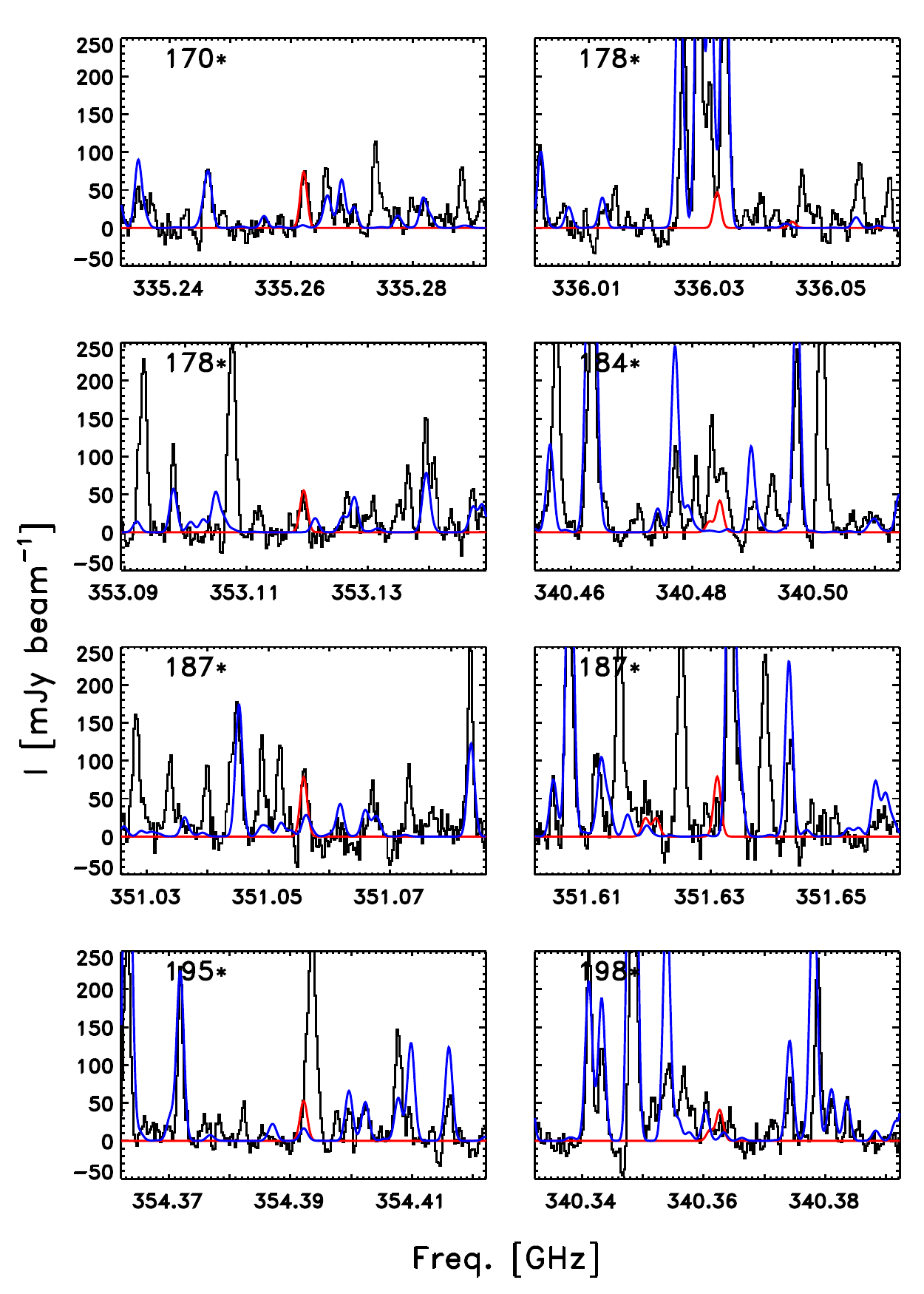}
      \label{sub:d-fullbeam_c}
                }
    \caption{Spectra for the 24 transitions of the symmetric conformer of CH$_3$OCHD$_2$ predicted to be the strongest given the derived excitation temperature and column density. In the panels, the red line shows the fit to the CH$_3$OCHD$_2$ transition while the blue lines show the predictions for of all other species identified as part of the PILS survey \citep[see, e.g.,][and references therein]{drozdovskaya19,manigand20}.}
    \label{fig:d-spectra_fullbeam}
  \end{center}
\end{figure*}

\begin{figure*}[ht]
  \begin{center}
    \subfloat[]{
      \includegraphics[width=0.8\columnwidth]{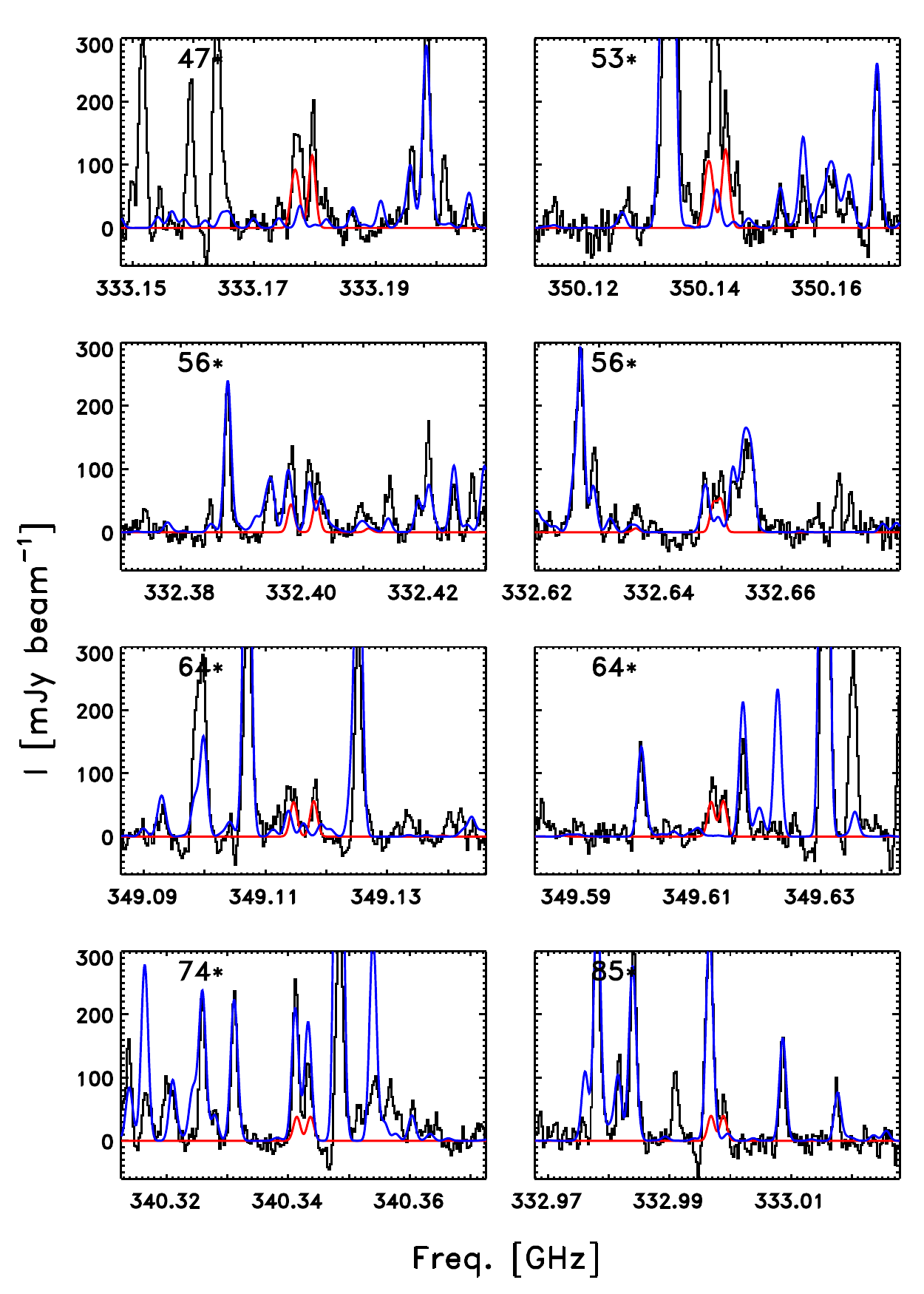}
      \label{sub:a-fullbeam_a}
                }
    \subfloat[]{
      \includegraphics[width=0.8\columnwidth]{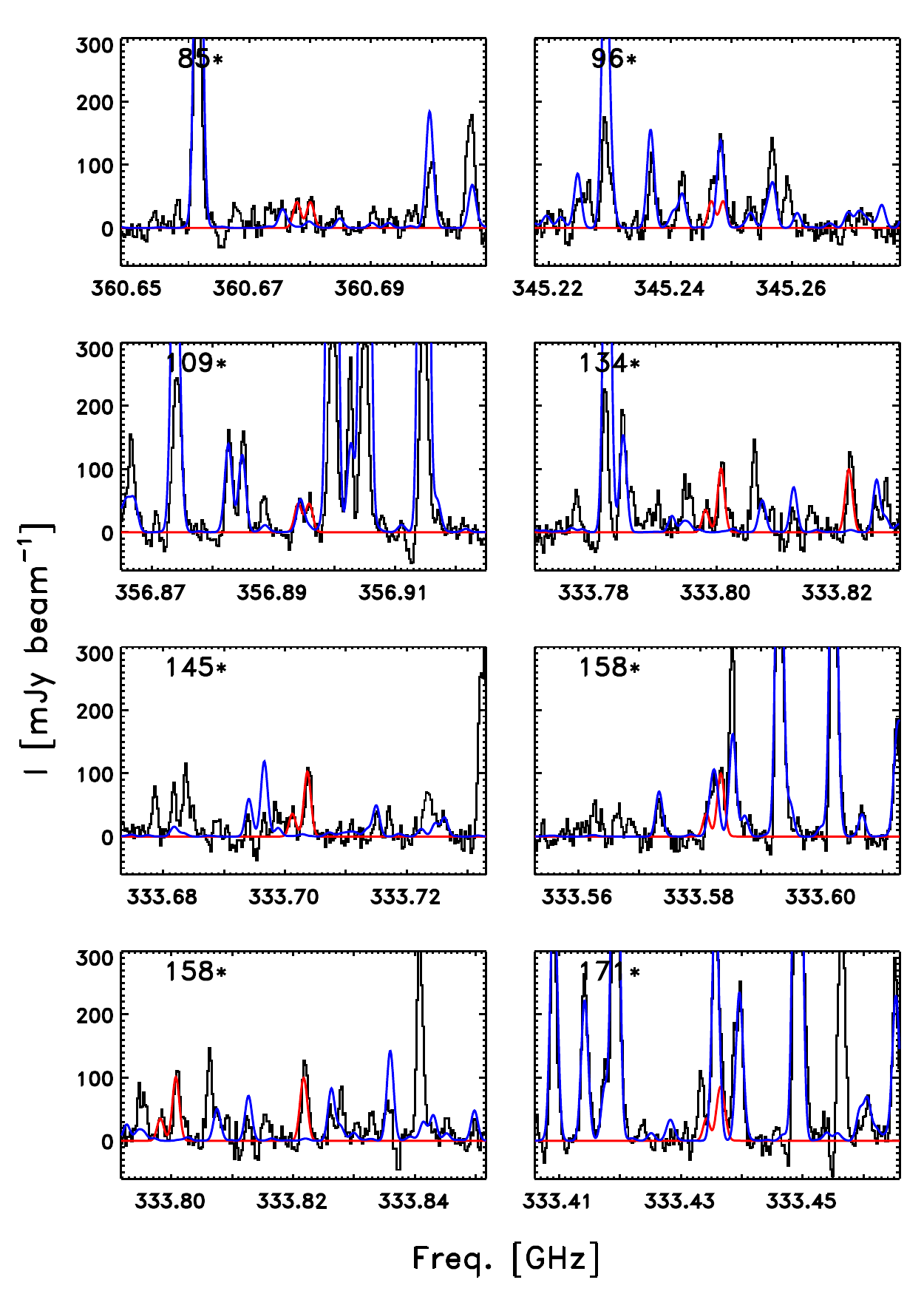}
      \label{sub:a-fullbeam_b}
                }
     \newline
   \subfloat[]{
      \includegraphics[width=0.8\columnwidth]{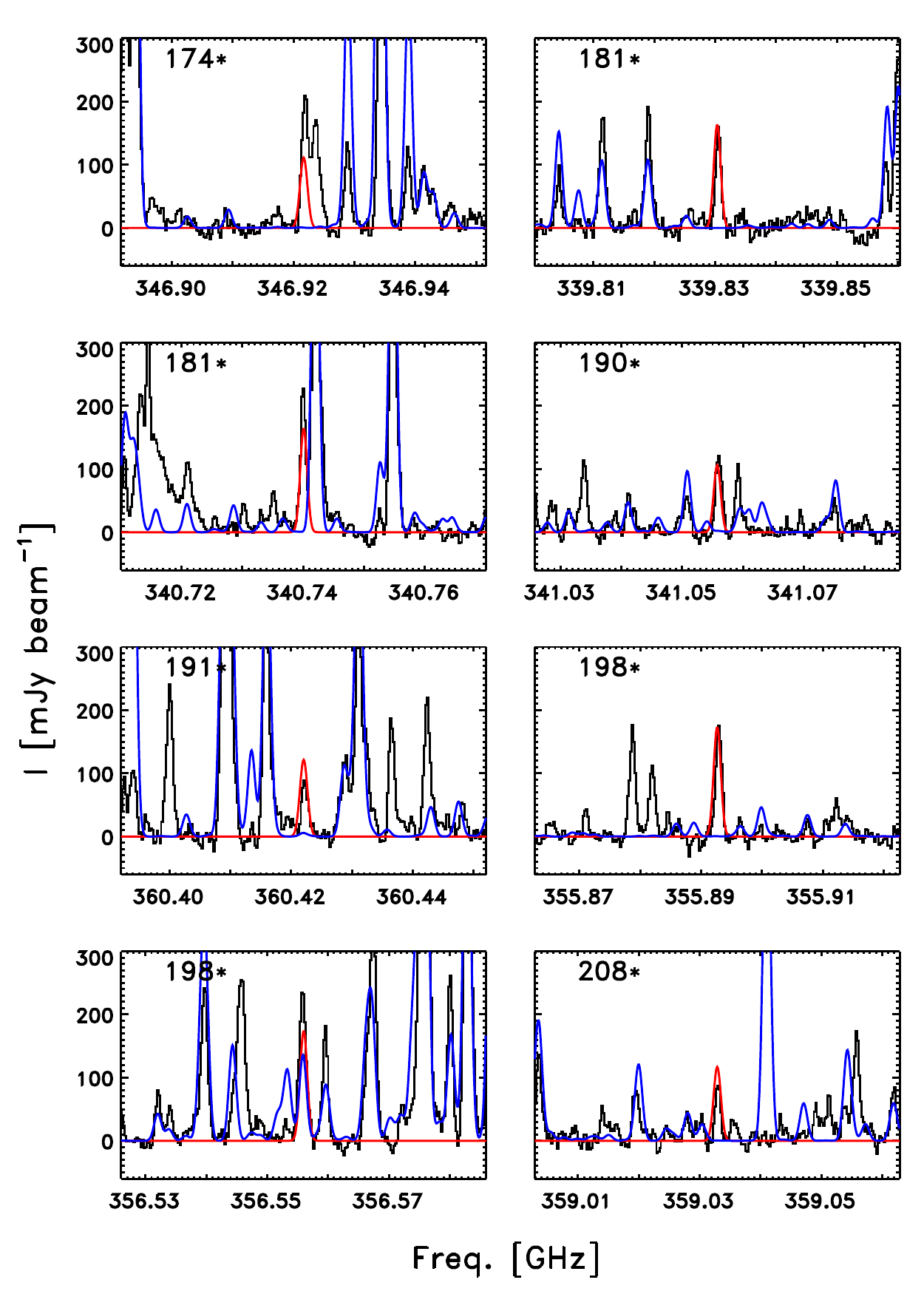}
      \label{sub:a-fullbeam_c}
                }
    \caption{Spectra for the 24 transitions of the asymmetric conformer of CH$_3$OCHD$_2$ predicted to be the strongest given the derived excitation temperature and column density.}
    \label{fig:a-spectra_fullbeam}
  \end{center}
\end{figure*}

Figs.~\ref{fig:d-spectra_fullbeam} and \ref{fig:a-spectra_fullbeam} show the 24 lines predicted in this manner to be the strongest for the asymmetic and symmetric conformers. For the two species good fits can be seen for about 20 transitions -- a number of which are well-isolated with peak intensities of up to about 200~mJy~beam$^{-1}$. The derived column densities are $6.1\times 10^{15}$~cm$^{-2}$ for the asymmetric and $2.7\times 10^{15}$~cm$^{-2}$ for the symmetric conformers. A'priori one would expect a factor 2 difference between the two due to the symmetry that occurs when two deuterium atoms replace two of the three hydrogen atoms in a CH$_3$-group of dimethyl ether. The factor 2.3 we derive when fitting the two conformers independently is consequently an additional verification of the assignment of the two conformers. 

Taken together the derived column densities imply a ratio between the column densities of the doubly-to-mono deuterated forms, $N$(CH$_3$OCHD$_2$)/$N$(CH$_3$OCH$_2$D) of 15--20\% for the two conformers. These imply D/H ratios that are about a factor of 5 higher than those inferred from the singly-to-non deuterated variants once corrected for the number of equivalent hydrogen atoms \citep[e.g., Appendix~B of][]{manigand19}. The trend of strong enhancements of the multi-deuterated variants also seen for other COMs thus also holds for DME, and provide additional constraints on its formation during the cold phases of star formation.

\section{Conclusion}
We have presented here a complete high-resolution study of the doubly-deuterated dimethylether over a decade in the frequency range 150~--1500~GHz (0.15--1.5 THz) in its two conformations, symmetic and asymmetric. We were able to determine an accurate set of spectroscopic parameters that allowed us to reproduce measurements with a standard deviation better than 80~kHz. Thanks to these frequency predictions, both conformers have been detected near the ``B'' component of the protostellar system IRAS 16293-2422. The doubly-deuterated species is found to be enhanced above the D/H ratio inferred from the mono-to-singly deuterated variants at a level similar to other organic molecules. These enhancements make CH$_3$OCH$_3$ an interesting target for further spectroscopic studies of other multiple deuterated variants, which in turn may shed light on the formation of this complex organic during the cold pre- and protostellar stages.

\begin{acknowledgements}
 This work has been supported by the “Programme National Physique et Chimie du Milieu Interstellaire” (PCMI) of CNRS/INSU with INC/INP co-funded by CEA and CNES. This project has received financial support from the CNRS through the MITI interdisciplinary programs. J.~C.~G. thanks the Centre National d’Etudes Spatiales (CNES) for a grant. The research of J.~K.~J is supported by the Independent Research Fund Denmark (grant number 0135-00123B).
\end{acknowledgements}

\bibliographystyle{aa} 
\bibliography{dme-2d}

\begin{thebibliography}{30}
\expandafter\ifx\csname natexlab\endcsname\relax\def\natexlab#1{#1}\fi

\bibitem[{Blukis {et~al.}(1963)Blukis, Kasai, \& Myers}]{blukis1963}
Blukis, U., Kasai, P.~H., \& Myers, R.~J. 1963, \jcp, 38, 2753

\bibitem[{Ceccarelli {et~al.}(2004)Ceccarelli, Johnstone, Adams, Lin, Neufeeld,
  \& Ostriker}]{ceccarelli2004}
Ceccarelli, C., Johnstone, D., Adams, F., {et~al.} 2004, in ASP Conf. Proc.,
  Vol. 323, 195

\bibitem[{{Drozdovskaya} {et~al.}(2019){Drozdovskaya}, {van Dishoeck}, {Rubin},
  {J{\o}rgensen}, \& {Altwegg}}]{drozdovskaya19}
{Drozdovskaya}, M.~N., {van Dishoeck}, E.~F., {Rubin}, M., {J{\o}rgensen},
  J.~K., \& {Altwegg}, K. 2019, \mnras, 490, 50

\bibitem[{Durig {et~al.}(1976)Durig, Li, \& Groner}]{durig1976}
Durig, J., Li, Y., \& Groner, P. 1976, J. Mol. Spectrosc., 62, 159

\bibitem[{Groner(1997)}]{groner1997}
Groner, P. 1997, \jcp, 107, 4483

\bibitem[{Groner(2012)}]{groner2012}
Groner, P. 2012, J. Mol. Spectrosc., 278, 52

\bibitem[{Groner {et~al.}(1986)Groner, Johnson, \& Durig}]{groner1986}
Groner, P., Johnson, R., \& Durig, J. 1986, J. Mol. Struct, 142, 363

\bibitem[{Ilyushin {et~al.}(2009)Ilyushin, Kryvda, \& Alekseev}]{ilyushin2009}
Ilyushin, V., Kryvda, A., \& Alekseev, E. 2009, J. Mol. Spectrosc., 255, 32

\bibitem[{{Jensen} {et~al.}(2021){Jensen}, {J{\o}rgensen}, {Kristensen},
  {Coutens}, {van Dishoeck}, {Furuya}, {Harsono}, \& {Persson}}]{jensen21}
{Jensen}, S.~S., {J{\o}rgensen}, J.~K., {Kristensen}, L.~E., {et~al.} 2021,
  arXiv e-prints, arXiv:2104.13411

\bibitem[{J{\o}rgensen {et~al.}(2016)J{\o}rgensen, Van~der Wiel, Coutens,
  Lykke, M{\"u}ller, Van~Dishoeck, Calcutt, Bjerkeli, Bourke, Drozdovskaya,
  {et~al.}}]{jorgensen16}
J{\o}rgensen, J., Van~der Wiel, M., Coutens, A., {et~al.} 2016, \aap, 595, A117

\bibitem[{{J{\o}rgensen} {et~al.}(2020){J{\o}rgensen}, {Belloche}, \&
  {Garrod}}]{jorgensen20}
{J{\o}rgensen}, J.~K., {Belloche}, A., \& {Garrod}, R.~T. 2020, \araa, 58, 727

\bibitem[{{J{\o}rgensen} {et~al.}(2018){J{\o}rgensen}, {M{\"u}ller}, {Calcutt},
  {Coutens}, {Drozdovskaya}, {{\"O}berg}, {Persson}, {Taquet}, {van Dishoeck},
  \& {Wampfler}}]{jorgensen18}
{J{\o}rgensen}, J.~K., {M{\"u}ller}, H.~S.~P., {Calcutt}, H., {et~al.} 2018,
  \aap, 620, A170

\bibitem[{Kisiel(2001)}]{kisiel2001}
Kisiel, Z. 2001, in Spectroscopy from Space (Springer), 91--106, see also
  PROSPE -- Programs for ROtational SPEctroscopy at
  \url{http://www.ifpan.edu.pl/~kisiel/prospe.htm}

\bibitem[{Linsky(2003)}]{linsky2003}
Linsky, J.~L. 2003, in Solar System History from Isotopic Signatures of
  Volatile Elements, ed. R.~Kallenbach, T.~Encrenaz, J.~Geiss, K.~Mauersberger,
  T.~Owen, \& F.~Robert (Springer), 49--60

\bibitem[{{Loinard} {et~al.}(2000){Loinard}, {Castets}, {Ceccarelli},
  {Tielens}, {Faure}, {Caux}, \& {Duvert}}]{loinard00}
{Loinard}, L., {Castets}, A., {Ceccarelli}, C., {et~al.} 2000, \aap, 359, 1169

\bibitem[{Lovas {et~al.}(1979)Lovas, Lutz, \& Dreizler}]{lovas1979}
Lovas, F., Lutz, H., \& Dreizler, H. 1979, J. Phys. Chem. Ref. Data, 8, 1051

\bibitem[{{Manigand} {et~al.}(2019){Manigand}, {Calcutt}, {J{\o}rgensen},
  {Taquet}, {M{\"u}ller}, {Coutens}, {Wampfler}, {Ligterink}, {Drozdovskaya},
  {Kristensen}, {van der Wiel}, \& {Bourke}}]{manigand19}
{Manigand}, S., {Calcutt}, H., {J{\o}rgensen}, J.~K., {et~al.} 2019, \aap, 623,
  A69

\bibitem[{{Manigand} {et~al.}(2020){Manigand}, {J{\o}rgensen}, {Calcutt},
  {M{\"u}ller}, {Ligterink}, {Coutens}, {Drozdovskaya}, {van Dishoeck}, \&
  {Wampfler}}]{manigand20}
{Manigand}, S., {J{\o}rgensen}, J.~K., {Calcutt}, H., {et~al.} 2020, \aap, 635,
  A48

\bibitem[{{Parise} {et~al.}(2006){Parise}, {Ceccarelli}, {Tielens}, {Castets},
  {Caux}, {Lefloch}, \& {Maret}}]{parise06deuterium}
{Parise}, B., {Ceccarelli}, C., {Tielens}, A.~G.~G.~M., {et~al.} 2006, \aap,
  453, 949

\bibitem[{{Parise} {et~al.}(2002){Parise}, {Ceccarelli}, {Tielens}, {Herbst},
  {Lefloch}, {Caux}, {Castets}, {Mukhopadhyay}, {Pagani}, \&
  {Loinard}}]{parise02}
{Parise}, B., {Ceccarelli}, C., {Tielens}, A.~G.~G.~M., {et~al.} 2002, \aap,
  393, L49

\bibitem[{{Parise} {et~al.}(2003){Parise}, {Simon}, {Caux}, {Dartois},
  {Ceccarelli}, {Rayner}, \& {Tielens}}]{parise03}
{Parise}, B., {Simon}, T., {Caux}, E., {et~al.} 2003, \aap, 410, 897

\bibitem[{{Persson} {et~al.}(2018){Persson}, {J{\o}rgensen}, {M{\"u}ller},
  {Coutens}, {van Dishoeck}, {Taquet}, {Calcutt}, {van der Wiel}, {Bourke}, \&
  {Wampfler}}]{persson18}
{Persson}, M.~V., {J{\o}rgensen}, J.~K., {M{\"u}ller}, H.~S.~P., {et~al.} 2018,
  \aap, 610, A54

\bibitem[{Pickett(1991)}]{pickett1991}
Pickett, H.~M. 1991, J. Mol. Spectrosc., 148, 371

\bibitem[{{Prodanovi{\'c}} {et~al.}(2010){Prodanovi{\'c}}, {Steigman}, \&
  {Fields}}]{prodanovic10}
{Prodanovi{\'c}}, T., {Steigman}, G., \& {Fields}, B.~D. 2010, \mnras, 406,
  1108

\bibitem[{Richard {et~al.}(2013)Richard, Margul{\`e}s, Caux, Kahane,
  Ceccarelli, Guillemin, Motiyenko, Vastel, \& Groner}]{richard2013}
Richard, C., Margul{\`e}s, L., Caux, E., {et~al.} 2013, \aap, 552, A117

\bibitem[{Smirnov {et~al.}(2014)Smirnov, Alekseev, Ilyushin, Margul{\'e}s,
  Motiyenko, \& Drouin}]{smirnov2014}
Smirnov, I., Alekseev, E., Ilyushin, V., {et~al.} 2014, J. Mol. Spectrosc.,
  295, 44

\bibitem[{Snyder {et~al.}(1974)Snyder, Buhl, Schwartz, Clark, Johnson, Lovas,
  \& Giguere}]{snyder1974}
Snyder, L., Buhl, D., Schwartz, P., {et~al.} 1974, \apj, 191, L79

\bibitem[{Townes \& Schawlow(1975)}]{townes1975}
Townes, C.~H. \& Schawlow, A.~L. 1975, Microwave spectroscopy, ed. N.~Y. D.~P.
  inc. (Courier Corporation)

\bibitem[{Yamamoto {et~al.}(2016)Yamamoto, Takamatsu, Hirano, \&
  Miura}]{yamamoto2016}
Yamamoto, C., Takamatsu, K., Hirano, K., \& Miura, M. 2016, J. Org. Chem., 81,
  7675

\bibitem[{Zakharenko {et~al.}(2015)Zakharenko, Motiyenko, Margul{\`{e}}s, \&
  Huet}]{zakharenko2015}
Zakharenko, O., Motiyenko, R.~A., Margul{\`{e}}s, L., \& Huet, T.~R. 2015, J.
  Mol. Spectrosc., 317, 41

\end{thebibliography}

\begin{appendix} 
\section{Experimental frequencies}
\begin{table*}[ht]
 \caption{Experimental frequencies measured in laboratory up to 1.5 THz for the symmetric conformer.}
\label{tab:exp_conf_s}
\begin{center}
\begin{tabular}{
	c
	c
	c
	c
	c
	c
	c
	S
	S
	S
	}
	\hline\hline
	\multicolumn{1}{c}{$\sigma$\tablefootmark{a}} & \multicolumn{1}{c}{$J'$} & \multicolumn{1}{c}{$K'_a$} & \multicolumn{1}{c}{$K'_c$} & \multicolumn{1}{c}{$J''$} & \multicolumn{1}{c}{$K''_a$} & \multicolumn{1}{c}{$K''_c$} & \multicolumn{1}{c}{Frequency} & \multicolumn{1}{c}{Uncertainty} & \multicolumn{1}{c}{o.-c.}\\
		 &  &  &  &  &  &  & \multicolumn{1}{c}{(MHz)} & \multicolumn{1}{c}{(MHz)} & \multicolumn{1}{c}{(MHz)} \\
		 \hline
$\cdots$ \\
1 & 13 & 4 &  9 & 13 & 3 & 10 & 154472.280 & 0.099 &  0.0540 \\
0 & 13 & 4 &  9 & 13 & 3 & 10 & 154473.864 & 0.099 &  0.0236 \\
1 & 20 & 3 & 18 & 20 & 2 & 19 & 154580.928 & 0.099 & -0.0216 \\
0 & 20 & 3 & 18 & 20 & 2 & 19 & 154582.476 & 0.099 & -0.0186 \\
1 & 12 & 4 &  8 & 12 & 3 &  9 & 155928.264 & 0.099 &  0.0032 \\
0 & 12 & 4 &  8 & 12 & 3 &  9 & 155929.956 & 0.099 &  0.0344 \\
1 & 11 & 4 &  7 & 11 & 3 &  8 & 157006.536 & 0.099 &  0.0627 \\
0 & 11 & 4 &  7 & 11 & 3 &  8 & 157008.156 & 0.099 &  0.0303 \\
1 & 10 & 4 &  6 & 10 & 3 &  7 & 157779.384 & 0.099 &  0.0408 \\
0 & 10 & 4 &  6 & 10 & 3 &  7 & 157780.896 & 0.099 &  0.0283 \\
0 & 23 & 5 & 19 & 22 & 6 & 16 & 157954.344 & 0.099 &  0.0286 \\
1 & 23 & 5 & 19 & 22 & 6 & 16 & 157955.208 & 0.099 &  0.0140 \\
1 &  9 & 4 &  5 &  9 & 3 &  6 & 158315.460 & 0.099 &  0.0267 \\
0 &  9 & 4 &  5 &  9 & 3 &  6 & 158316.576 & 0.099 &  0.0247 \\
0 &  8 & 4 &  4 &  8 & 3 &  5 & 158674.920 & 0.099 & -0.0426 \\
$\cdots$ \\

\hline
\end{tabular}
\tablefoot{\tablefoottext{a}{Symmetry number: 0(A), 1(E).}}
\end{center}
\end{table*}

\begin{table*}[ht]
 \caption{Experimental frequencies measured in laboratory up to 1.5 THz for the asymmetric conformer.}
\label{tab:exp_conf_a}
\begin{center}
\begin{tabular}{
	c
	c
	c
	c
	c
	c
	c
	S
	S
	S
	}
	\hline\hline
	\multicolumn{1}{c}{$\sigma$\tablefootmark{a}} & \multicolumn{1}{c}{$J'$} & \multicolumn{1}{c}{$K'_a$} & \multicolumn{1}{c}{$K'_c$} & \multicolumn{1}{c}{$J''$} & \multicolumn{1}{c}{$K''_a$} & \multicolumn{1}{c}{$K''_c$} & \multicolumn{1}{c}{Frequency} & \multicolumn{1}{c}{Uncertainty} & \multicolumn{1}{c}{o.-c.}\\
		 &  &  &  &  &  &  & \multicolumn{1}{c}{(MHz)} & \multicolumn{1}{c}{(MHz)} & \multicolumn{1}{c}{(MHz)} \\
		 \hline
$\cdots$ \\
1 & 28 &  3 & 25 & 28 & 2 & 26 & 150936.099 & 0.050 &    0.0201 \\
0 & 28 &  3 & 25 & 28 & 2 & 26 & 150937.072 & 0.050 &   -0.0012 \\
1 & 17 &  3 & 15 & 17 & 2 & 16 & 152080.811 & 0.050 &    0.0429 \\
0 & 17 &  3 & 15 & 17 & 2 & 16 & 152082.764 & 0.050 &    0.0257 \\
1 & 33 &  4 & 29 & 33 & 3 & 30 & 152137.688 & 0.050 &    0.0841 \\
1 & 23 &  2 & 21 & 23 & 1 & 22 & 152153.268 & 0.050 &   -0.0116 \\
0 & 23 &  2 & 21 & 23 & 1 & 22 & 152154.840 & 0.050 &   -0.0291 \\
0 & 20 &  3 & 18 & 19 & 4 & 15 & 152639.222 & 0.050 &   -0.0005 \\
1 & 20 &  3 & 18 & 19 & 4 & 15 & 152640.886 & 0.050 &   -0.0189 \\
1 & 20 &  4 & 16 & 20 & 3 & 17 & 153413.952 & 0.099 &    0.0247 \\
0 & 20 &  4 & 16 & 20 & 3 & 17 & 153415.500 & 0.099 &    0.0270 \\
1 & 17 &  2 & 16 & 17 & 1 & 17 & 155133.960 & 0.099 &    0.0193 \\
0 & 17 &  2 & 16 & 17 & 1 & 17 & 155136.336 & 0.099 &    0.0180 \\
0 & 10 &  0 & 10 &  9 & 1 &  9 & 155368.536 & 0.099 &    0.0722 \\
1 & 10 &  0 & 10 &  9 & 1 &  9 & 155368.536 & 0.099 &   -0.1006 \\
$\cdots$ \\

\hline
\end{tabular}
\tablefoot{\tablefoottext{a}{Symmetry number: 0(A), 1(E).}}
\end{center}
\end{table*}

\section{Predicted frequencies}
\begin{table*}[ht]
 \caption{Predicted transition frequencies of doubly-deuterated DME in the ground-vibrational state for the symmetric conformer.}
\label{tab:pred_conf_s}
\begin{center}
\begin{tabular}{
	c
	c
	c
	c
	c
	c
	c
	S
	S
	S
	S
	S
	}
	\hline\hline
	\multicolumn{1}{c}{$\sigma$\tablefootmark{a}} & \multicolumn{1}{c}{$J'$} & \multicolumn{1}{c}{$K'_a$} & \multicolumn{1}{c}{$K'_c$} & \multicolumn{1}{c}{$J''$} & \multicolumn{1}{c}{$K''_a$} & \multicolumn{1}{c}{$K''_c$} & \multicolumn{1}{c}{Frequency} & \multicolumn{1}{c}{Uncertainty} & \multicolumn{1}{c}{Spin} & & \multicolumn{1}{c}{$E$}\\
		 &  &  &  &  &  &  & \multicolumn{1}{c}{(MHz)} & \multicolumn{1}{c}{(MHz)} & \multicolumn{1}{c}{Weight} & \multicolumn{1}{c}{S} & \multicolumn{1}{c}{(cm$^{-1}$)} \\
\hline
1 & 15  &  4 &  11 & 15 &  3 & 12 &  150168.493 & 0.004 & 4 & 7.9007 &   82.305 \\
 0 & 15  &  4 &  11 & 15 &  3 & 12 &  150169.934 & 0.005 & 4 & 7.9010 &   82.305 \\
 0 & 28  & 15 &  14 & 29 & 14 & 15 &  150188.926 & 0.010 & 4 & 0.0775 &  407.265 \\
 0 & 28  & 15 &  13 & 29 & 14 & 16 &  150188.926 & 0.010 & 4 & 0.0775 &  407.265 \\
 1 & 28  & 15 &  14 & 29 & 14 & 16 &  150189.636 & 0.011 & 4 & 0.0775 &  407.265 \\
 1 & 28  & 15 &  13 & 29 & 14 & 15 &  150189.708 & 0.009 & 4 & 0.0775 &  407.265 \\
 0 & 27  &  2 &  26 & 26 &  3 & 23 &  150530.744 & 0.018 & 4 & 0.1022 &  220.054 \\
 1 & 27  &  2 &  26 & 26 &  3 & 23 &  150533.598 & 0.018 & 4 & 0.1022 &  220.054 \\
 1 &  5  &  2 &   4 &  4 &  1 &  3 &  150807.089 & 0.002 & 4 & 4.0124 &   11.780 \\
 0 &  5  &  2 &   4 &  4 &  1 &  3 &  150808.179 & 0.003 & 4 & 4.0125 &   11.780 \\
 0 & 30  &  5 &  25 & 30 &  4 & 26 &  151255.992 & 0.009 & 4 & 2.7557 &  291.944 \\
 1 & 30  &  5 &  25 & 30 &  4 & 26 &  151256.456 & 0.007 & 4 & 2.7557 &  291.944 \\
 1 & 15  & 10 &   6 & 16 &  9 &  8 &  151518.542 & 0.008 & 4 & 0.3512 &  145.838 \\
 0 & 15  & 10 &   6 & 16 &  9 &  7 &  151519.578 & 0.009 & 4 & 0.3512 &  145.838 \\
 0 & 15  & 10 &   5 & 16 &  9 &  8 &  151519.578 & 0.009 & 4 & 0.3512 &  145.838 \\
$\cdots$ \\

\hline
\end{tabular}
\tablefoot{\tablefoottext{a}{Symmetry number: 0(A), 1(E).}}
\end{center}
\end{table*}

\begin{table*}[ht]
 \caption{Predicted transition frequencies of doubly-deuterated DME in the ground-vibrational state for the asymmetric conformer.}
\label{tab:pred_conf_a}
\begin{center}
\begin{tabular}{
	c
	c
	c
	c
	c
	c
	c
	S
	S
	S
	S
	S
	}
	\hline\hline
	\multicolumn{1}{c}{$\sigma$\tablefootmark{a}} & \multicolumn{1}{c}{$J'$} & \multicolumn{1}{c}{$K'_a$} & \multicolumn{1}{c}{$K'_c$} & \multicolumn{1}{c}{$J''$} & \multicolumn{1}{c}{$K''_a$} & \multicolumn{1}{c}{$K''_c$} & \multicolumn{1}{c}{Frequency} & \multicolumn{1}{c}{Uncertainty} & \multicolumn{1}{c}{Spin} & & \multicolumn{1}{c}{$E$}\\
		 &  &  &  &  &  &  & \multicolumn{1}{c}{(MHz)} & \multicolumn{1}{c}{(MHz)} & \multicolumn{1}{c}{Weight} & \multicolumn{1}{c}{S} & \multicolumn{1}{c}{(cm$^{-1}$)} \\
\hline
 1 &  28 &   3 & 25 & 28 & 2 & 26 & 150936.079  &  0.005  & 4  & 3.5816 & 241.633 \\
 0 &  28 &   3 & 25 & 28 & 2 & 26 & 150937.073  &  0.005  & 4  & 3.5817 & 241.633 \\
 0 &  25 &   5 & 20 & 24 & 6 & 19 & 151482.629  &  0.004  & 4  & 1.0671 & 206.034 \\
 1 &  25 &   5 & 20 & 24 & 6 & 19 & 151483.750  &  0.004  & 4  & 1.0669 & 206.034 \\
 1 &  17 &   3 & 15 & 17 & 2 & 16 & 152080.768  &  0.003  & 4  & 7.3428 &  94.479 \\
 0 &  17 &   3 & 15 & 17 & 2 & 16 & 152082.738  &  0.003  & 4  & 7.3430 &  94.479 \\
 1 &  33 &   4 & 29 & 33 & 3 & 30 & 152137.604  &  0.004  & 4  & 2.0711 & 335.886 \\
 0 &  33 &   4 & 29 & 33 & 3 & 30 & 152137.941  &  0.005  & 4  & 2.0711 & 335.886 \\
 1 &  23 &   2 & 21 & 23 & 1 & 22 & 152153.280  &  0.006  & 4  & 4.8360 & 162.649 \\
 0 &  23 &   2 & 21 & 23 & 1 & 22 & 152154.869  &  0.006  & 4  & 4.8361 & 162.649 \\
 0 &  20 &   3 & 18 & 19 & 4 & 15 & 152639.222  &  0.004  & 4  & 1.7455 & 126.757 \\
 1 &  20 &   3 & 18 & 19 & 4 & 15 & 152640.905  &  0.004  & 4  & 1.7455 & 126.757 \\
 1 &  20 &   4 & 16 & 20 & 3 & 17 & 153413.927  &  0.003  & 4  & 7.5813 & 133.093 \\
 0 &  20 &   4 & 16 & 20 & 3 & 17 & 153415.473  &  0.004  & 4  & 7.5814 & 133.093 \\
 1 &  11 &   4 &  8 & 12 & 1 & 11 & 153421.035  &  0.007  & 4  & 0.0136 &  51.116 \\
$\cdots$ \\

\hline
\end{tabular}
\tablefoot{\tablefoottext{a}{Symmetry number: 0(A), 1(E).}}
\end{center}
\end{table*}

\end{appendix}

\end{document}